\documentclass[preprint]{elsarticle}

\usepackage{hyperref}
\usepackage{mathtools}

\journal{arXiv.org}

\begin{document}

\begin{frontmatter}

\title{Variable-spot ion beam figuring}

\author{Lixiang Wu}
\ead{wulx@mail.ustc.edu.cn}

\author{Keqiang Qiu}
\ead{blueleaf@ustc.edu.cn}

\author{Shaojun Fu}
\ead{sjfu@ustc.edu.cn}

\address{National Synchrotron Radiation Laboratory, University of Science and Technology of China, Hefei 230029, China}

\begin{abstract}
This paper introduces a new scheme of ion beam figuring (IBF), or rather variable-spot IBF, which is conducted at a constant scanning velocity with variable-spot ion beam collimated by a variable diaphragm.
It aims at improving the reachability and adaptation of the figuring process within the limits of machine dynamics by varying the ion beam spot size instead of the scanning velocity.
In contrast to the dwell time algorithm in the conventional IBF, the variable-spot IBF adopts a new algorithm, which consists of the scan path programming and the trajectory optimization using pattern search.
In this algorithm, instead of the dwell time, a new concept, integral etching time, is proposed to interpret the process of variable-spot IBF.
We conducted simulations to verify its feasibility and practicality.
The simulation results indicate the variable-spot IBF is a promising alternative to the conventional approach.
\end{abstract}


\end{frontmatter}


\section{Introduction}

IBF is a powerful technique for enabling ultra-precision finishing of optical surfaces \cite{arnold_ultra-precision_2010,shore_ultra-precision_2012}.
As a deterministic method, it has proven effective and applicable in processing both large optics \cite{guo_large_2012,ghigo_ion_2014} and small optical components \cite{drueding_ion_1995}.
For years, it has seen a growing demand of the high-quality ultra-precision optics finished by IBF technologies in many fields such as astronomy telescopes \cite{guo_large_2012,ghigo_ion_2014}, synchrotron beamlines \cite{peverini_ion_2010,siewert_metrology_2013,alcock_characterization_2015}, and the extreme ultraviolet lithography \cite{weiser_ion_2009}.

As is well known, the conventional figuring process can be considered as a typical convolution process, in which the controlled ion beam with fixed Gaussian-shaped profile scans over a workpiece with variable dwell velocities.
The calculation of dwell times that determine the figuring process is an ill-posed deconvolution problem meanwhile it plays a crucial role in achieving high accuracy of the IBF \cite{drueding_contouring_1995,jiao_algorithm_2009}.
And this method can be traced back to the computer-controlled optical surfacing (CCOS), which is a breakthrough technology in the optical fabrication field \cite{jones_rapid_1990}.

We believe the conventional method is not the only scheme and here put forward a new one, variable-spot IBF, with some practical benefits.
Instead of varying the scanning velocity of ion beam, our method is realized by dynamically adjusting the spot size of ion beam via a variable diaphragm.
And the variable-spot IBF would prefer to run at a constant scanning velocity.
In contrast to the conventional approach, where a fixed diaphragm is usually used \cite{schindler_ion_2001} and thus an invariant removal function is obtained, the variable-spot IBF involves a variable diaphragm that collimates the ion beam and then forms a spatially and temporally variable ion beam spot on the workpiece surface, which results in a variable-spot removal function.
We can really benefit a lot from the two key properties of variable-spot IBF, i.e., variable spot and constant scanning velocity.
The ion beam with variable spot has a better adaptation to the target material removal because ion beams with different spot sizes are applicable in removing surface errors of different spatial frequencies.
Moreover, it is more convenient and economical to adjust the variable diaphragm than to control the scanning motion of ion source.

The variable-spot IBF is not a convolution process so we cannot adopt the dwell time algorithm of the conventional IBF.
Recently, we gave a new interpretation of the process of the variable-spot figuring process based on the concept of integral etching time and the spatial-temporal analysis, and then proposed an algorithm for the variable-spot IBF.
The algorithm consists of two parts:  scan path programming and trajectory optimization.  By the scan path programming, the two-dimensional (2D) figuring is reduced into the one-dimensional (1D) figuring.
Then the trajectory optimization of 1D figuring using pattern search is conducted on the basis of the S-curve transformation with consideration of the limits of machine dynamics.
The optimization can directly generate the motion profile, which gives the variable-spot method an obvious advantage over the conventional method that requires an approximate conversion from dwell times to dwell velocities.

\section{Variable-spot ion beam}

\subsection{Design of variable diaphragm}

\begin{figure}
\centering\includegraphics[width=.4\columnwidth]{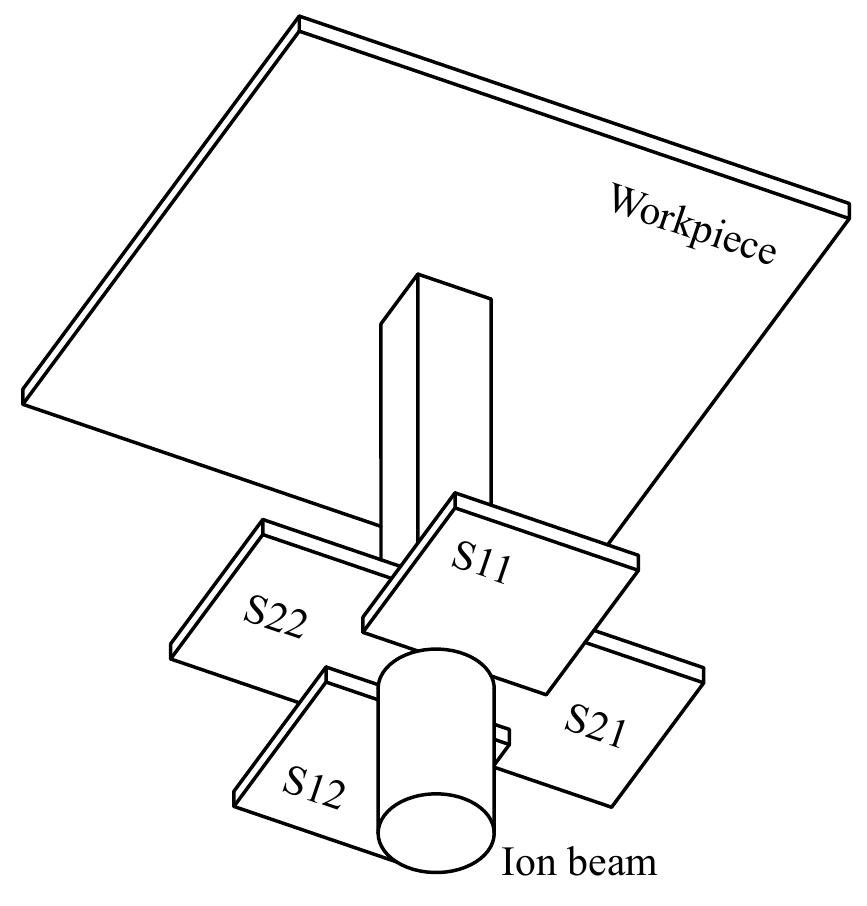}
\caption{IBF with the variable diaphragm. Note that S11(S21) and S12(S22) is a pair of shutters, which is driven by a single linear motor.}
\label{fig:IBF-with-diaphragm}
\end{figure}

We have designed a variable diaphragm for dynamically collimating the ion beam emitted from a broad ion source and adjusting its dimensions.
As is illustrated in Figure~\ref{fig:IBF-with-diaphragm}, the variable diaphragm with two pairs of shutters collimates the round ion beam into a rectangular beam in the course of IBF.
The variable diaphragm is dragged by a linear motor and its aperture diameter $\Phi(t)$ is proportional to the translation distance $D(t)$ of the motor such that
\begin{equation}
\Phi(t) = kD(t),
\end{equation}
where $k$ is the gear ratio. Note that only one pair of shutters moves along the scan path and the other is temporarily fixed during the figuring process.
The aperture diameter $\Phi(t)$ hereafter refers to the distance of the two moving shutters.
Note also that the center of ion beam always coincides with the aperture center during the IBF process, thus the beam profile is symmetric.

\subsection{Assumptions of variable-spot removal function}

The Kaufman ion source is widely used in IBF \cite{kaufman_technology_1982} and it typically induces a removal function with Gaussian profile in the figuring of planar optics \cite{liao_mathematical_2014-1}.
The normalized Gaussian removal function $B_n(x_1,x_2;t)$ is assumed as
\begin{equation} \label{eq:NormalizedBeamFun}
B_n(x_1, x_2;t) = \frac{\Phi_1(t)\Phi_2}{\Phi_0^2} \exp{\left\{-4\ln{2}\left[(\frac{x_1-s_1(t)}{\Phi_1(t)})^2 + (\frac{x_2-s_2(t)}{\Phi_2})^2\right]\right\}},
\end{equation}
where $s_1(t)$ and $s_2(t)$ indicate the center position of the beam spot at time $t$, $\Phi_1(t)$ and $\Phi_2$ represent the aperture dimensions of the variable diaphragm, and $\Phi_0$ is the limit of aperture diameter.
Note that the defined removal function varies along the $x_1$ direction in the course of IBF and its width or full width at half maximum (FWHM), $\Phi_2$, in the $x_2$ direction is specified before the figuring process.

\begin{figure}
\centering\includegraphics[width=.7\columnwidth]{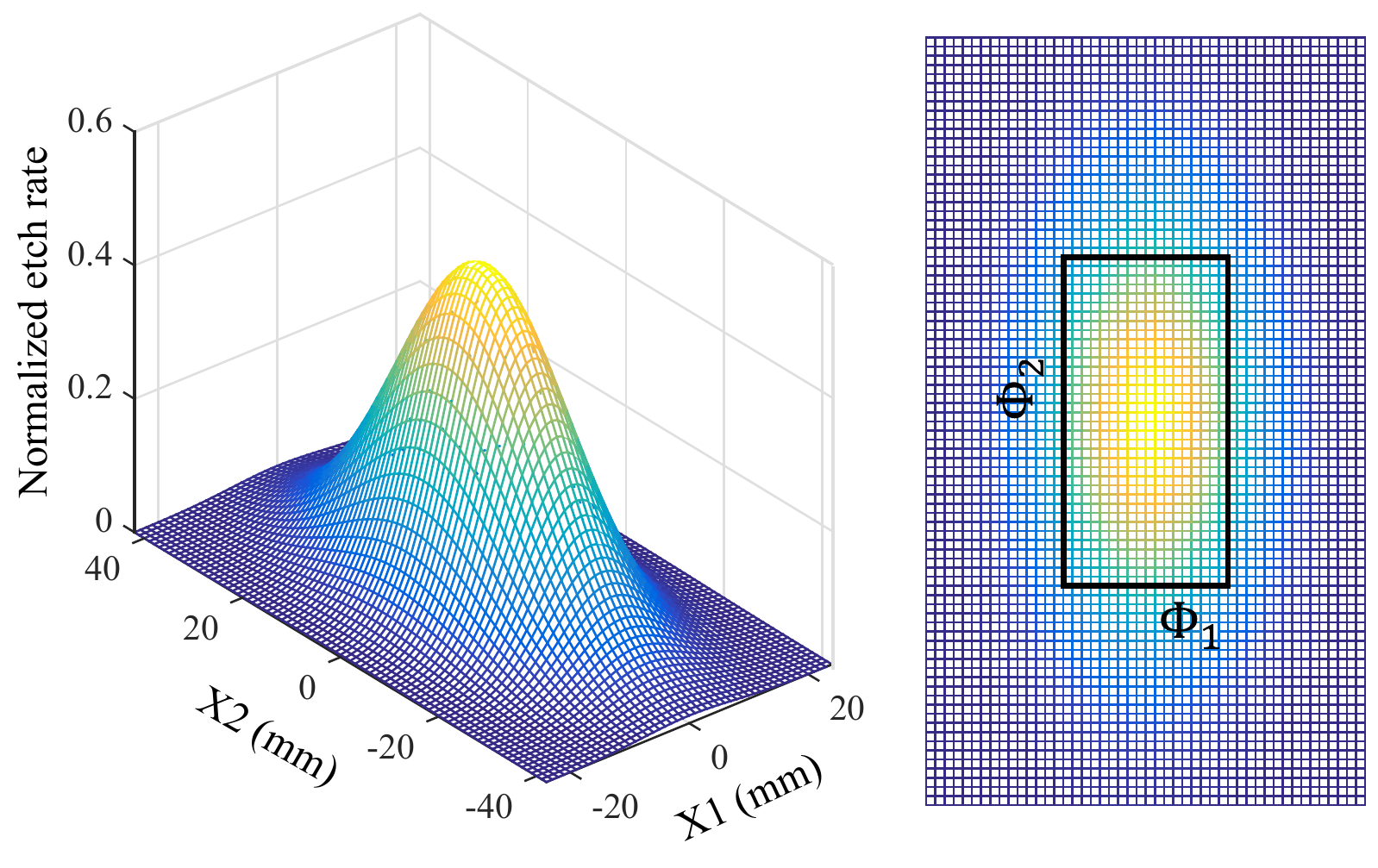}
\caption{Normalized Gaussian removal function. The rectangle with two sides of lengths of $\Phi_1$ and $\Phi_2$ on the right indicates the diaphragm aperture, where $\Phi_1$ is 18~mm and $\Phi_2$ is 36~mm.}
\label{fig:removal-functions}
\end{figure}

For convenience, we make an assumption that the FWHMs of removal function equals to the aperture dimensions of the variable diaphragm and the peak of the removal function linearly decreases as the diaphragm aperture narrows.
The details are illustrated in Figure~\ref{fig:removal-functions}, where $\Phi_1$ and $\Phi_2$ represent the aperture dimensions of the variable diaphragm and equal to the FWHMs of the Gaussian removal function along the $x_1$-axis and $x_2$-axis.
Herein, $\Phi_1$ is specified as the gap distance between the pair between moving shutters, that is, S11 and S12 (see Figure~\ref{fig:IBF-with-diaphragm}), and $\Phi_2$ represents the gap distance of another pair or the pair of temporarily fixed shutters.

\subsection{Practical considerations}

It is reasonable to make assumptions of the variable-spot removal function, which serve as a convenience to the simulations that demonstrate the process of variable-spot IBF and verify the feasibility of the proposed method.
However, these assumptions cannot be applicable in the real cases of variable-spot IBF since the removal functions should be accurately measured by experiments or the mathematical modeling based on material removal characteristics.
W. Liao \textit{et al.} studied the mathematical modeling and application of removal functions during deterministic IBF of optical surfaces \cite{liao_mathematical_2014-1,liao_mathematical_2014-2}.
It suggests that mathematical models that are developed theoretically and verified experimentally can help to build the removal function database.
In practice, we should also develop a database of variable-spot removal functions, which is applicable in the variable-spot IBF of optical elements.

\section{Spatio-temporal analysis of 1D figuring}

It has been mentioned that the 2D figuring can be reduced into the 1D figuring by the scan path programming.
Here, we introduce a general approach based on the integral etching time to analyze the 1D figuring process, which is the basis of the real world 2D figuring.

\subsection{Integral etching time}
\label{sec:etching-time}

The normalized removal function defined in Equation~\ref{eq:NormalizedBeamFun} is reduced into the 1D function
\begin{equation} \label{eq:NormalizedBeamFun1D}
B_n(s;t) = \frac{\Phi_1(t)\Phi_2}{\Phi_0^2} \exp{\left[-4\ln{2}\left(\frac{s-s_1(t)}{\Phi_1(t)}\right)^2\right]}.
\end{equation}
And the integral etching time $T_{etch}(s)$ is defined as
\begin{equation} \label{eq:IntEtchTime}
T_{etch}(s) = \int_{t=0}^{\infty} B_n(s;t) dt.
\end{equation}
The new term, integral etching time, is introduced to describe the integrated time weighted by the normalized removal function, which is abbreviated as etching time hereafter.
So the material removal $R(s)$ can be reinterpreted as
\begin{equation} \label{eq:RemovalFun}
R(s) = \int_{t=0}^{\infty} B(s;t) dt
     = \int_{t=0}^{\infty} B_n(s;t) r_{etch,max} dt
	 = T_{etch}(s) r_{etch,max},
\end{equation}
where $B(s;t)$ is the Gaussian removal function and $r_{etch,max}$ is the maximum etch rate or the peak of the etch rate profile.

\subsection{Spatio-temporal visualization}

\begin{figure}
\centering\includegraphics[width=.65\columnwidth]{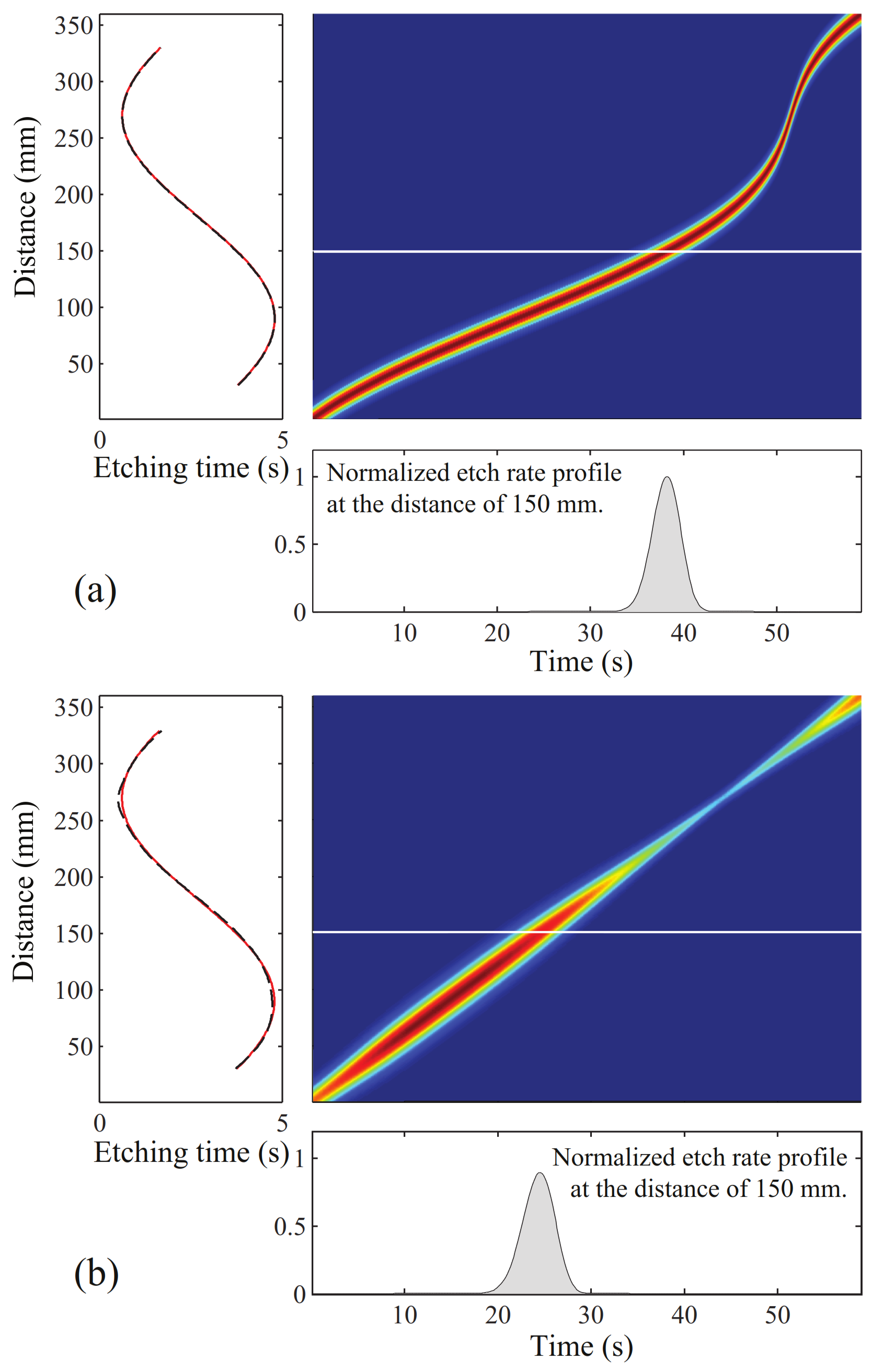}
\caption{Spatio-temporal visualization of (a) the conventional IBF process (b) and the variable-spot IBF process. The sinusoidal solid curves indicate the target material removals and the dotted curves are the calculated results.}
\label{fig:velocity-vs-diaphragm}
\end{figure}

The spatio-temporal visualization of two different 1D figuring processes is presented to further explore the relationship between the variable-spot IBF and the conventional IBF.
Figure~\ref{fig:velocity-vs-diaphragm}~(a) illustrates a conventional figuring process with a time-invariant Gaussian removal function in a spatio-temporal coordinate system, where the twisted ``rainbow" indicates the track of the normalized Gaussian removal function.
Figure~\ref{fig:velocity-vs-diaphragm}~(b) demonstrates a typical variable-spot IBF process, which operates at a constant scanning velocity with a variable-spot ion beam.
Both cases shown in Figure~\ref{fig:velocity-vs-diaphragm} have an identical target material removal (the sinusoidal curves on the left).
The former result is calculated using the \textit{deconvlucy} function in Matlab, which is an implementation of the Lucy-Richardson method \cite{biggs_acceleration_1997}.
The latter result is obtained by the trajectory optimization based on the modified fine adjustment algorithm \cite{wu_algorithms_2015}.
The preliminary simulations indicate that variable-spot IBF is a possible alternative technique to the conventional IBF.

\section{Algorithm for variable-spot IBF}

\subsection{Scan path programming}

\begin{figure}
\centering\includegraphics[width=.8\columnwidth]{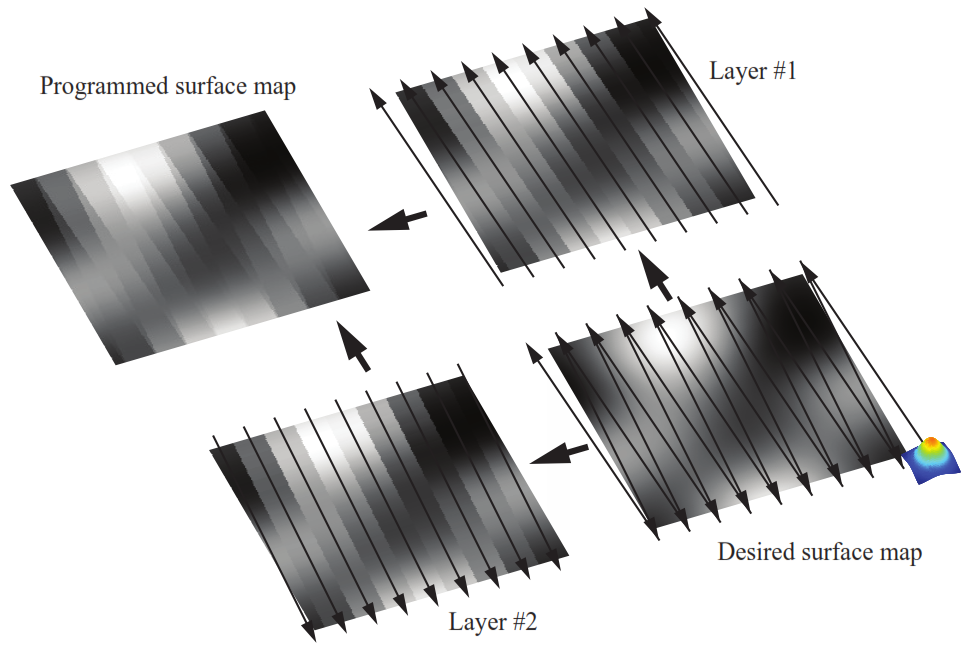}
\caption{Illustration of scan-path programming. There are three processing steps, i.e., rastering, grouping, and superposing, from the desired surface map to the programmed surface map.}
\label{fig:scan-path}
\end{figure}

We proposed a simple scheme for programming scan paths of the variable-spot ion beam.
It consists of three steps: rastering, grouping, and superposition, which is shown in Figure~\ref{fig:scan-path}.
First, the ion beam scans boustrophedonically over the desired surface map or etch-depth map along a zig-zag track.
Then, the zig-zag track is broken down into many linear paths, which are classified as several groups, and correspondingly the averaged surface maps are reduced into “belt-like surfaces”.
And finally, the programmed surface map is obtained by point-to-point superposition of the belt-like surfaces.
In the step of grouping, the linear paths that have both an identical direction and a uniform transverse distance of the width of removal function, which equals to $\Phi_2$, are classified as a layer.
The number of linear paths within a distance of $\Phi_2$ should be positive even such as $2, 4, 6,$ and so on, which determines the number of layers in the grouping step.
Note that after the process of scan path programming, we can horizontally rotate the workpiece to a quarter turn, then conduct the scan path programming again, and the crossing superposition of two times of programming may result in a more smoothing programmed map.

After the programming of scan paths, programmed etching time profiles, corresponding to profiles of the belt-like surfaces, are calculated according to Equation~\ref{eq:RemovalFun}.
The etching time profile is the ratio of the programmed etch-depth profile, i.e. the profile of belt-like surface in the programmed surface map, to the maximum etch rate.

\subsection{Trajectory optimization}

The beam scanning velocity should be optimized first.
Then the objective is to find a sequence of feasible motion trajectories of the motor driving the pair of moving shutters that make the calculated etching time profiles as close as possible to the programmed etching time profiles.

\subsubsection{Beam scanning velocity}

The initial estimate of beam scanning velocity is
\begin{equation} \label{eq:BSV}
v_{scan} = \frac{r \frac{\Phi_2}{\Phi_0} \int_{-\infty}^{\infty} \exp[-4\ln{2}(\frac{x}{\Phi_0})^2] dx}{\max(T_{etch})}.
\end{equation}
where $r$ is an empirical parameter with default value of $1$ and $\max(T_{etch})$ indicates the maximum of the target etching times.

\subsubsection{S-curve transformation}

In our previous work \cite{wu_algorithms_2015}, we first found that there exists a correspondence relationship between the trapezoidal velocity profile of the driving motor and the S-shaped etching time profile along the scan path.
The shape of an etching time profile can be adjusted by tuning control parameters of the corresponding trapezoidal velocity profile, which is the key point of S-curve transformation.
Moreover, the programmed etching time profiles can be represented by B-splines \cite{jester_b-spline_2011}, and theoretically a B-spline can be divided into a group of S-curves taking the stationary points as the division points.
So, for a desired etching time profile, the trajectory optimization is to find a sequence of trapezoidal velocity profiles determining an etching time profile close to the desired etching time profile.
The root-mean-square deviation (RMSD) of the calculated etching time profile and the desired etching time profile should be within the specified maximum deviation.

Trapezoidal velocity profiles \cite{jeon_generalized_2000} are applied to providing smooth motion for motors of the variable diaphragm.
Currently, we only use trapezoidal velocity profiles to define the motion trajectory of motors.
Because the trapezoidal velocity profile is the simplest scheme for motion smoothing and the number of its control parameters is also less than the other motion profiles.

\subsubsection{Trajectory optimization using pattern search}

Pattern search is a family of direct-search methods without using derivatives of the problem to be optimized, which is deterministic and suitable for optimizing the problem of one-dimension variable-spot IBF.
For this optimization problem, it follows the basic generalized pattern search algorithm of tree steps: initialization, search and poll steps, and parameters update \cite{audet_analysis_2002}.

The first step is to determine the initial or starting point and the objective function.
The starting point, which is a vector or pattern, includes the perturbations of diaphragm diameter at the division points, the acceleration and deceleration for adjusting the trapezoidal velocity profile.
The objective function is the RMSD of the calculated etching time curve and the target etching time spline.
And then perform the search and possibly the poll steps.
Finally, update parameters iteratively until adapting to the target.
Based on the S-curve transformation, the parameters that determines the trapezoidal velocity profiles are iteratively adjusted for improving the fit to the S-shaped etching time profiles which comprise the desired etching time spline.

\section{Simulations and discussions}

Simulations have been conducted in two different cases.
The first simulation demonstrates the detailed processes of the aforementioned trajectory optimization in the case of 1D figuring.
The second simulates a real world process of 2D figuring and aims at verifying the feasibility of variable-spot IBF.
The desired surface map is a random rough surface with Gaussian statistics generated by the RSG code \cite{david_surface_2015}.

\subsection{Trajectory optimization of 1D figuring}

The simulation parameters and their values are shown in Table~\ref{tab:SimParams}, where $v_u$ and $a_u$ represent the upper limits of velocity and acceleration, $N_{iter,max}$ indicates the maximum number of iterations and is specified as the stop condition.
Other parameters have been illustrated above.

\begin{table}
\centering\caption{Simulation parameters.} \label{tab:SimParams}
\begin{tabular}{cl|cl}
\hline
Parameter & Value & Parameter & Value \\
\hline
$v_{scan}$ & 1.10~mm/s & $k$ & 1/1000 \\
$v_u$ & 600~steps/s & $a_u$ & 30~steps/s$^2$ \\
$\Phi_0$ & 36~mm & $r$ & 0.9 \\
$N_{iter,max}$ & 240 & $r_{etch,max}$ & 1~nm/s\\ 
\hline
\end{tabular}
\end{table}

\begin{figure}
\centering\includegraphics[width=.75\columnwidth]{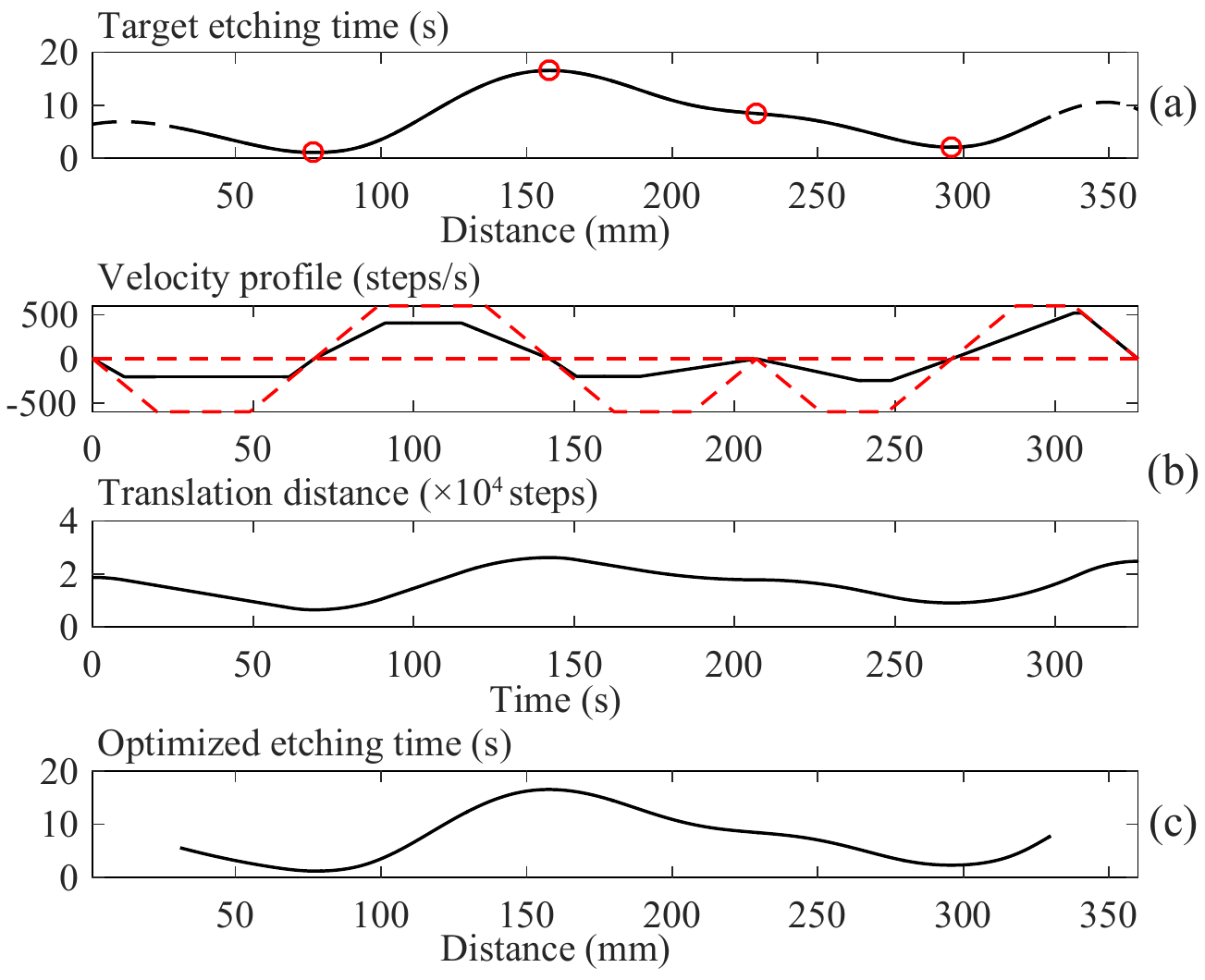}
\caption{The processing results of an iteration in trajectory optimization, including (a) the target etching time spline, (b) the motion profile, and (c) the optimized etching time curve. First, the target spline is divided into 5 S-curves, then the trapezoidal velocity profile is adjusted within the feasible regions (marked by the dotted lines), and finally, an optimized etching time curve is obtained by performing the conversions shown in Figure~\ref{fig:CalEtchTime}.}
\label{fig:ResOfIter}
\end{figure}

Figure~\ref{fig:ResOfIter} shows the processing results of an iteration in the trajectory optimization.
First, the target etching time spline is divided into a group of S-curves taking the stationary points as the division points (the circles marked in Figure~\ref{fig:ResOfIter}~(a)).
And then it polls a pattern to adjust the motion profile of the variable diaphragm.
The pattern is a vector containing the control parameters of trapezoidal velocity profiles.
The feasible regions, marked by the dotted lines in Figure~\ref{fig:ResOfIter}~(b), indicate the limits of machine dynamics of the variable diaphragm.
The parameter optimization of trapezoidal velocity profiles is conducted within the feasible regions.
Finally, the optimized etching time is calculated according to the aforementioned spatio-temporal visualization.
The objective is to minimize the return value of the objective function, the RMSD of the target etching time profile and the optimized etching time profile, until meeting the stop condition.

\begin{figure}
\centering\includegraphics[width=.75\columnwidth]{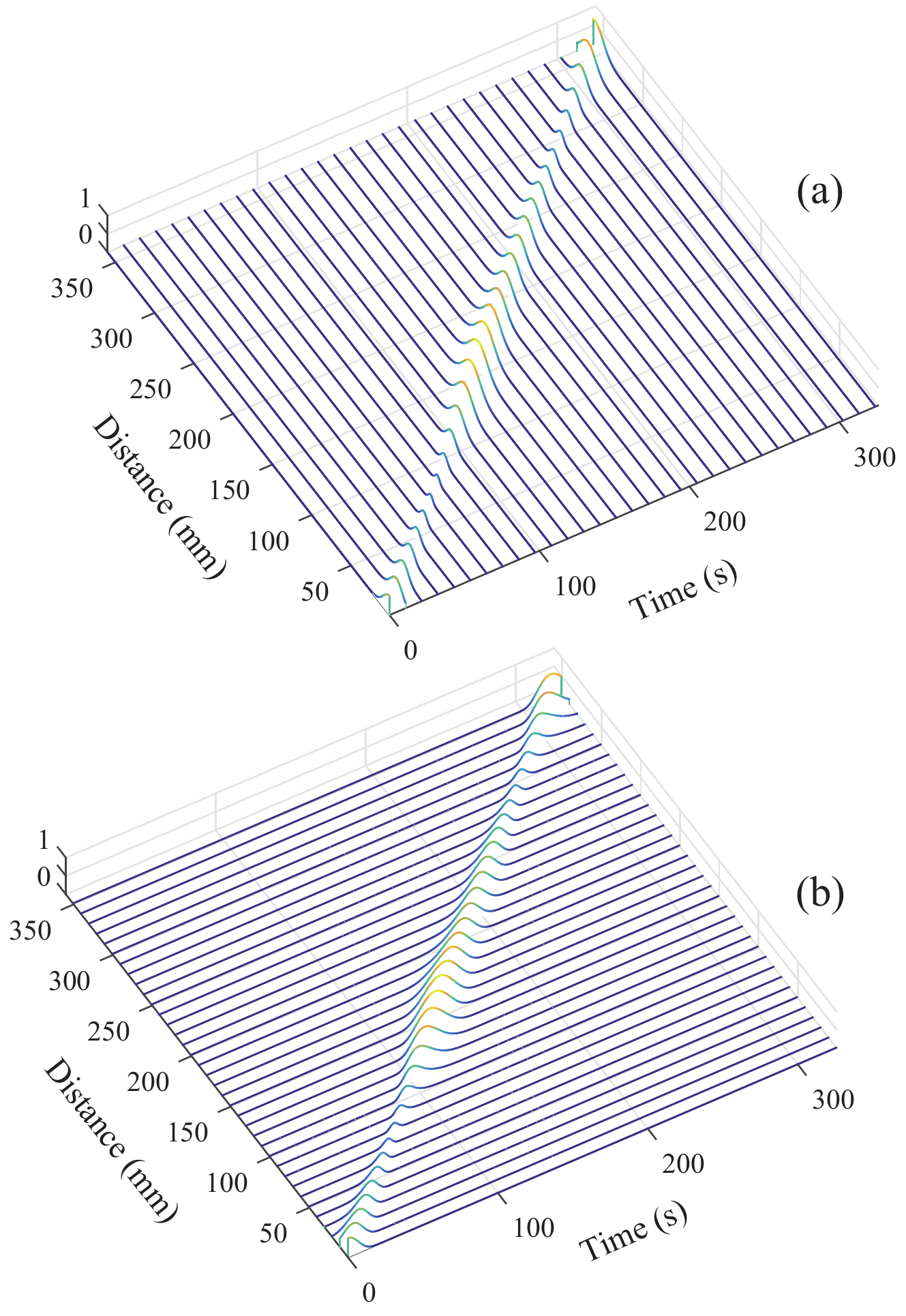}
\caption{Calculation of optimized etching times. (a) Mapping the scanning track by normalized variable-spot removal functions and (b) obtaining etching time profiles by grid-based interpolation are two basic steps for calculating the optimized etching times, which are integrals of the etching time profiles.}
\label{fig:CalEtchTime}
\end{figure}

As shown in Figure~\ref{fig:CalEtchTime}, there contains two basic steps to calculate the optimized etching times.
First, the figuring process is represented by mapping the scanning track of the variable beam spot by normalized 1D removal functions according to the polled pattern (Figure~\ref{fig:CalEtchTime}~(a)).
Then the etching time curves are obtained by grid-based interpolation of the results of the previous step (Figure~\ref{fig:CalEtchTime}~(b)), and the optimized etching times are calculated by numerically integrating the interpolated curves along the time axis.

\subsection{2D figuring with variable-spot ion beam}

\begin{figure}
\centering\includegraphics[width=.75\columnwidth]{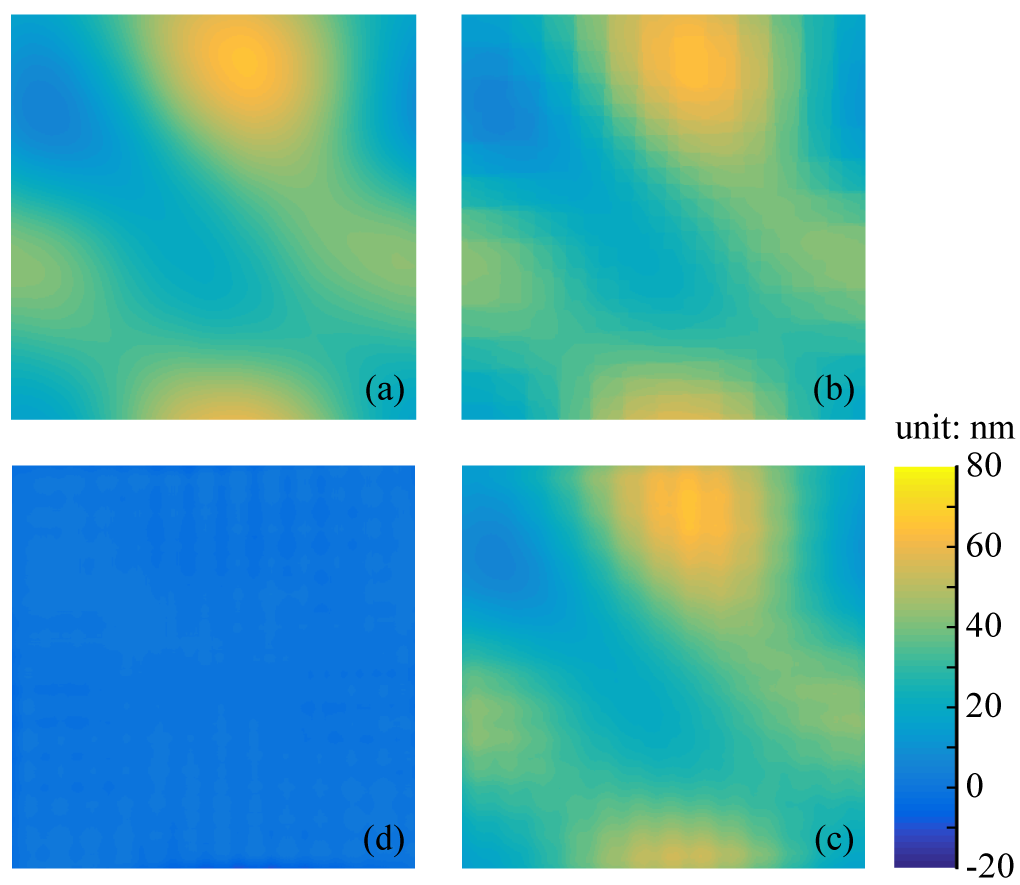}
\caption{(a) The desired surface map, (b) the programmed surface map, (c) the predicted surface map, and (d) the error map. Note that the programmed surface map is a crossing result and the error map presents etch-depth differences between the desired surface map and the predicted surface map.}
\label{fig:error-map}
\end{figure}

Figure~\ref{fig:error-map} shows the desired surface map, the programmed surface map, the predicted surface map, and the error map that indicates the differences between the desired map and the predicted map.
The programmed surface map is a crossing result obtained by conducting two times of scan path programming.
After the scan path programming, we get programmed etching time profiles and then perform trajectory optimization on the programmed etching time profiles one by one, which can be considered as performing many times of trajectory optimization of 1D figuring.
The predicted surface map is the simulated result of variable-spot IBF according to the programmed scan path and the optimized trajectory.

\begin{figure}
\centering\includegraphics[width=.5\columnwidth]{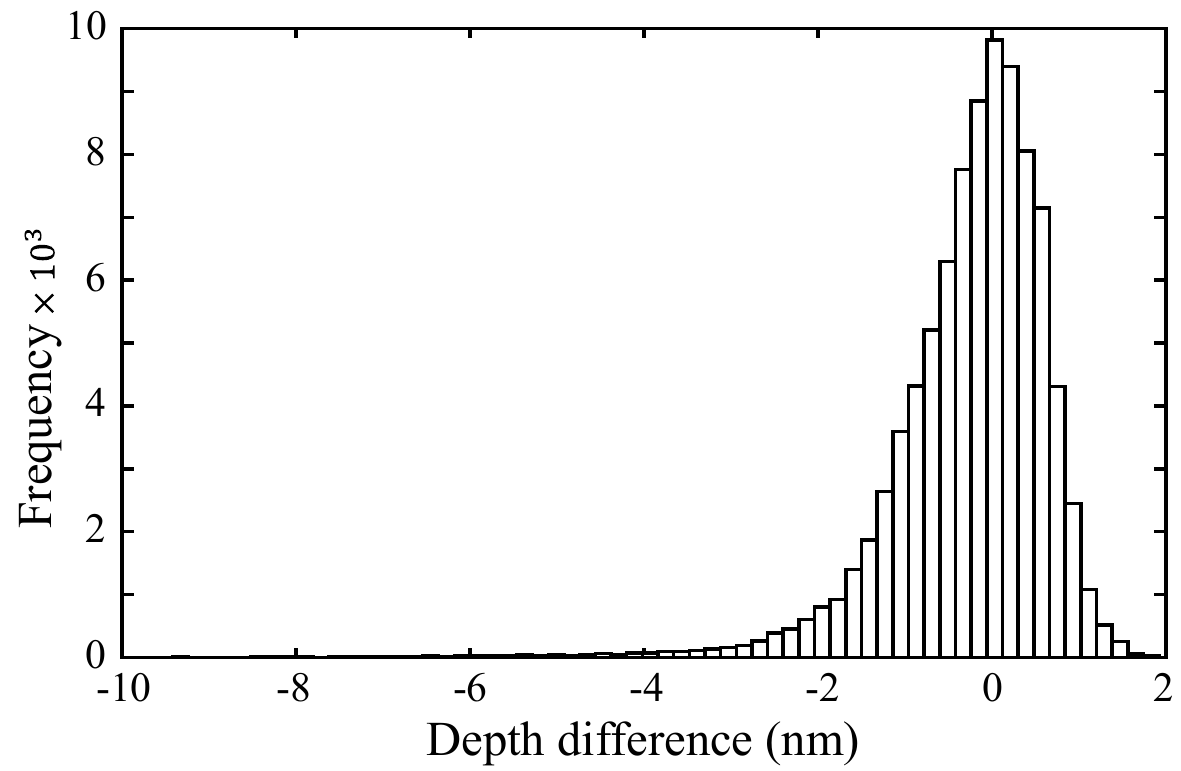}
\caption{Distribution of etch-depth differences of the desired surface map and the predicted surface map.}
\label{fig:dist-profile}
\end{figure}

The PV value and RMS height of the desired surface map are 60~nm and 12.9~nm.
After the figuring in a single iteration, as is summarized in Figure~\ref{fig:dist-profile}, the PV value decreases to about 11.5~nm and the RMS height becomes about 0.9~nm.
And in the centering area of 240~mm$\times$240~mm, the PV value and the RMS height are about 4.9~nm and 0.7~nm.
It indicates that variable-spot IBF is competent for the precision fabrication of optical surfaces.
Note that, in this simulation, the dimensions of these surface map are 300~mm$\times$300~mm and the width of removal function is selected as 24~mm.
Under this condition, the RMS height of about 0.9~nm is obtained.

\subsubsection{Parameter optimization in scan path programming}

\begin{figure}
\centering\includegraphics[width=.7\columnwidth]{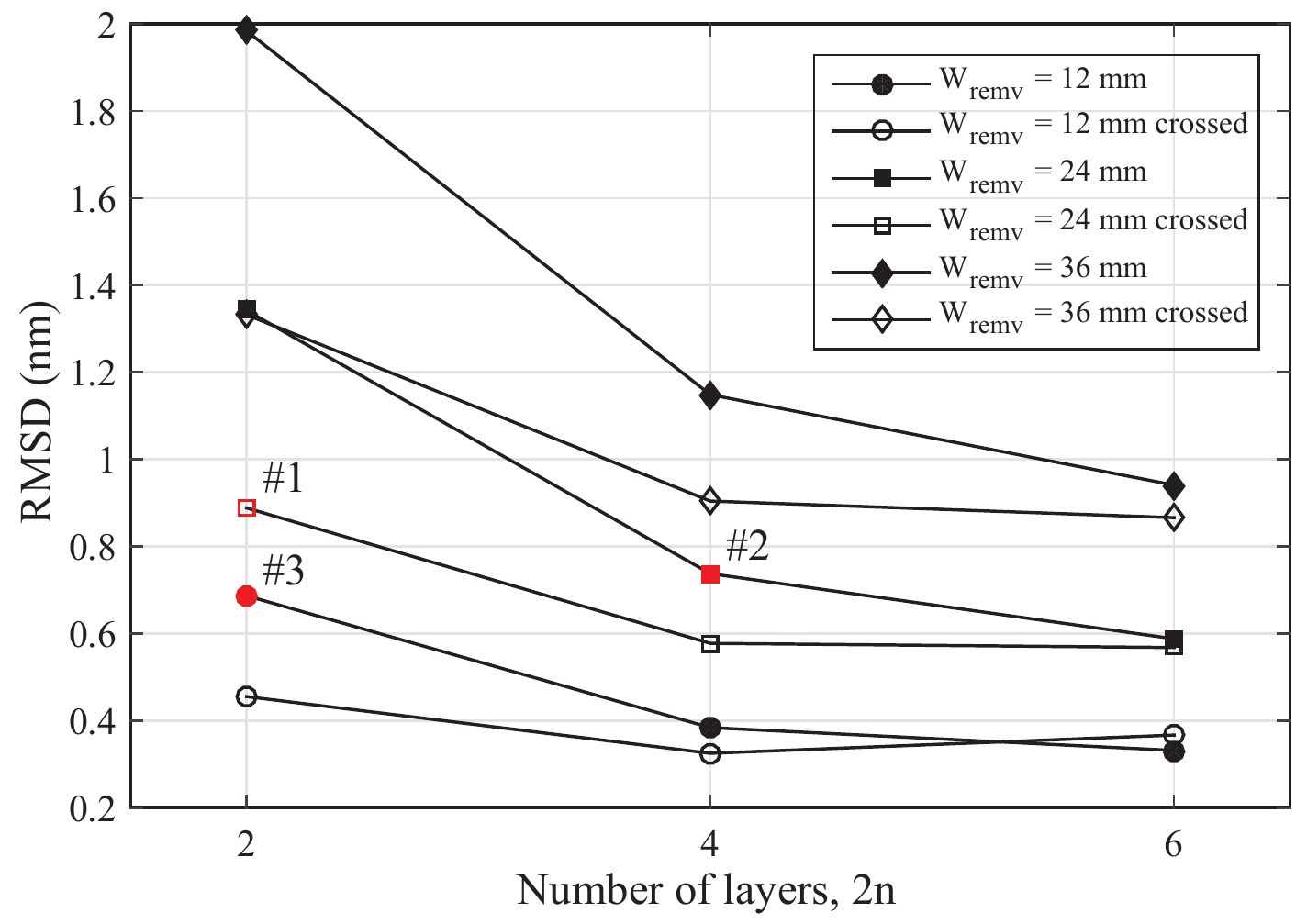}
\caption{The variation of RMSD with the width of removal function and the number of layers. Note that the RMSD crossed refers to the RMSD of the target and the crossing result of two times of scan-path programming.}
\label{fig:scan-opt}
\end{figure}

In the scan path programing, the 2D optical surface is divided into several groups of belt-like surfaces, which is a process of dimensional reduction that approximately transforms the 2D figuring process into a series of 1D figuring processes.
The objective of scan path programming is to search a group of reasonable parameters, including the width of removal function and the number of layers, under which the RMSD of the desired surface map and the programmed surface map would be no greater than the desired value.
Figure~\ref{fig:scan-opt} shows that the RMSD, of the desired surface map and the programmed surface map, decreases as the width of removal function, $W_{remv}$, narrows and the number of layers, $2n$, increases.
Note that the width of removal function, $W_{remv}$, equals to the gap distance, $\Phi_2$, of the pair of temporarily fixed shutters according to the assumption of variable-spot removal function.

\begin{figure}
\centering\includegraphics[width=.6\columnwidth]{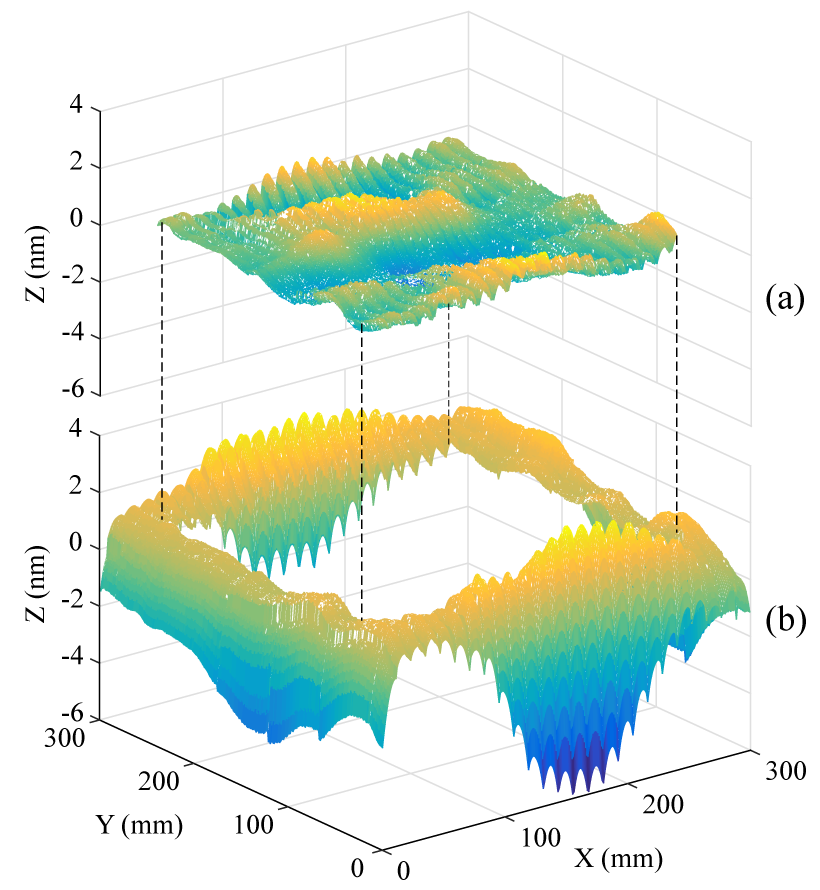}
\caption{(a) Centering area and (b) margin area of the error map in case \#2.}
\label{fig:error-map-2}
\end{figure}

In addition to the figuring performance, the cost-efficiency should be considered.
The narrower width and the more layers, to some extent, would lead to the smaller RMSD, or rather the more smoothing surface map.
Meanwhile, it would take more working time and that is not economical.
Table~\ref{tab:ParamCmp} presents the predicted results after the figuring process in three different cases, of which the specified parameters are marked in Figure~\ref{fig:scan-opt}.
The case \#1 have been described in this section.
In the case \#2, the number of layers is 4 and the width of removal function is 24~mm.
In the case \#3, the number of layers is 2 while the width of removal function decrease into 12~mm.
They are nearly equal in the total length of scan path for those cases.
However, the case \#3 requires 791.2~min, which is almost 3.6 times longer than what either of the other two cases takes.
Moreover, the case \#2 has a better performance compared with the case \#1 and the RMS height and PV in the centering area of 240~mm$\times$240~mm are decreased into 0.40~nm and 2.78~nm respectively after the figuring (see Figure~\ref{fig:error-map-2}).

To summarize the above, the case \#2 is the best choice among the three cases and it indicates that the width of removal function and the number of layers should be balanced in the scan path programming and the crossing of two times of scan-path programming is not the most effective approach to improve the figuring performance.
Moreover, as is illustrated in Figure~\ref{fig:error-map-2}, the surface roughness of the margin area is significantly greater than that of the centering area, which indicates the preallocated margins are important for improving the figuring performance.

\begin{table}
\centering\caption{Comparison of simulated results.} \label{tab:ParamCmp}
\begin{tabular}{l|c|c|c}
\hline
 & case \#1 & case \#2 & case \#3 \\
\hline
$2n$, $W_{remv}$ & 2, 24~mm (crossed) & 4, 24~mm & 2, 12~mm \\
\hline
RMS/PV (nm) &  &  &  \\
in whole area & 0.91/11.47 & 0.74/9.17 & 1.54/19.04  \\
in centering area & 0.66/4.91 & 0.40/2.78 & 1.24/7.44 \\
in margin area & 1.23/11.42 & 1.10/9.17 & 1.98/19.04 \\
\hline
Total time (min) & 220.6 & 221.0 & 791.2 \\
\hline
\end{tabular}
\end{table}

\subsubsection{Fine tuning of the maximum etch rate}

\begin{figure}
\centering\includegraphics[width=.6\columnwidth]{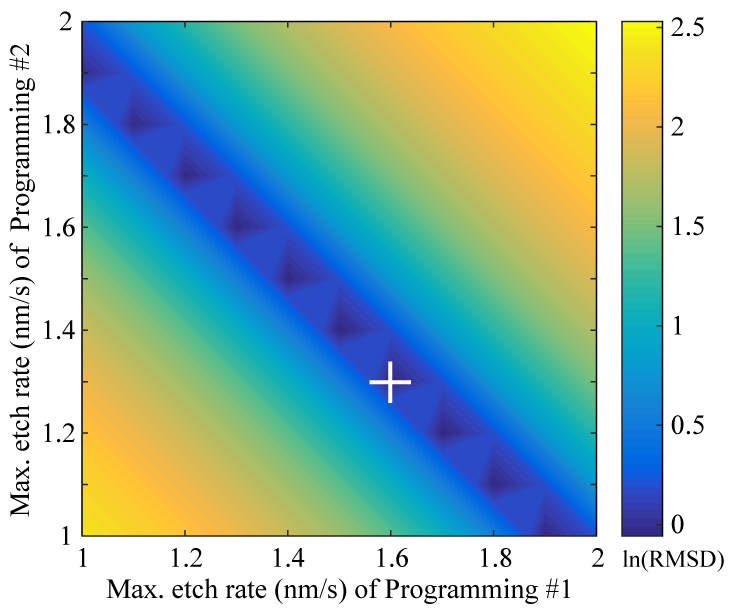}
\caption{Fine tuning of maximum etch rates. The programming \#1 and \#2 indicate the first and the second programmed scan path.}
\label{fig:fine-tune}
\end{figure}

It has been found that the predicted surface map usually has an overall-shift deviation with the desired surface map.
Our strategy for decreasing the overall-shift deviation is finely tuning the maximum etch rate by varying the distance between the ion source and the workpiece.
As is shown in Figure~\ref{fig:fine-tune}, the deviation, or rather the RMSD, varies as the maximum etch rates changes.
There are many “lowlands” around the line defined by $x + y = 2.9$.
The best point is marked by a cross in Figure~\ref{fig:fine-tune}, where the RMSD is about 0.9~nm while the two maximum etch rates are 1.6~nm/s and 1.3~nm/s, respectively.
Note that there are two times of scan path programming so we can finely tune the maximum etch rates for the two programmed scan paths.
If there is no crossing, we can finely tune the maximum etch rate for only one time.

\section{Conclusion}

The concepts of figuring with variable spots at constant scanning speed are introduced and the algorithm for variable-spot IBF has also been briefly described.
Two simulations are conducted.
For the first simulation, its result indicates that the trajectory optimization using pattern search based on S-curve transformation is feasible.
And the result of the second simulation has verified the feasibility of variable-spot IBF applied to the precision figuring of optical surfaces.
The variable-spot IBF is a promising alternative to the conventional IBF.
The preliminary analysis suggests that we would benefit from the combination of the two techniques of IBF.
Also, the idea of surface figuring or polishing with variable spots is a generalized concept for optical fabrication, from which other advanced techniques of optical fabrication such as plasma jet machining, magnetorheological finishing, and fluid jet polishing can also benefit.

\section*{Acknowledgement}

The work was partially supported by the National Natural Science Foundation of China under Grant No. 11275201.

\section*{References}

\bibliography{ref}

\end{document}